\documentclass[a4paper, 10pt]{article}
\usepackage{graphicx}

\usepackage{alltt}

\usepackage{amsmath,amsthm,amscd,amssymb,mathtools,mathrsfs,bm,dsfont,latexsym}

\usepackage{yhmath} 
\usepackage{srctex} 
\usepackage[dvipsnames]{color}
\usepackage{tabularx}
\usepackage{multicol}
\usepackage{wrapfig}
\usepackage{sidecap}
\usepackage{multirow}

\usepackage[ruled]{algorithm2e}
\newcommand{\w}{\ensuremath{\mathbin{\wedge}}} 

\newcommand{\cl}[2]{\ensuremath{\mathit{Cl}_{#1,#2}}}

\newcommand{\bbR}{\ensuremath{\mathbb{R}}}

\newcommand{\bbC}{\ensuremath{\mathbb{C}}}
\newcommand{\bbH}{\ensuremath{\mathbb{H}}}

\newcommand{\ii}{\mathbf{i}}
\newcommand{\jj}{\mathbf{j}}
\newcommand{\kk}{\mathbf{k}}

\def\m#1{\mathsf{#1}}

\def\e#1{\mathbf{e}_{#1}} 

\newcommand{\bq}{\ensuremath{\mathbf{q}}}
\newcommand{\bp}{\ensuremath{\mathbf{p}}}

\newcommand{\bv}{\ensuremath{\mathbf{v}}}

\newcommand{\bV}{\ensuremath{\bm{V}}}

\def\m#1{\mathsf{#1}}

\usepackage[T1]{fontenc}

\begin{document}

\begin{center}
\LARGE
\textbf{Square roots of real and  complex quaternions}\\[6pt]
\small
\textbf{Adolfas Dargys$^{*}$, Art{\=u}ras Acus$^{**}$}\\[6pt]
$^{*}$Center for Physical Sciences and Technology, Semiconductor
Physics Institute, \\ Saul{\.e}tekio 3, LT-10257
Vilnius, Lithuania\\
adolfas.dargys@gmail.lt\\
$^{**}$Institute of Theoretical Physics and Astronomy, Vilnius
University,\\ Saul{\.e}tekio 3, LT-10257 Vilnius, Lithuania\\
arturas.acus@tfai.vu.lt\\[6pt]

\end{center}

 \begin{abstract} Square roots of real and complex (complexified)  quaternions, namely, the Hamilton quaternion, coquaternion,
 nectorine, and conectorine are investigated. The isomorphism between the  quaternions and
 multivectors in Clifford algebras  is employed for this purpose. Root examples for all named quaternions are
  presented from which follows that the real and complex quaternionic roots may assume multiple discrete or  continuous  forms,
 or there may be no roots at all. Examples are presented.\\

 \vskip 2mm


\end{abstract}

\tableofcontents

\section{Introduction}\label{sec:1}

Quaternions are an indispensable tool in finding  optimal trajectories in robotics and space flights as well as in a general control of
rotational motion and interpolation of rotations \cite{Ward1997,Altmann2005,Turner2006}. Also, the quaternions find application in classical and
quantum physics~\cite{Kravchenko2003,Adler1994}, or even in the cosmology~\cite{Hamilton2023}. A number of practical books have been published
already~\cite{Gurlebeck1997,Dorst07}. There are four kinds of non-commutative quaternions~\cite{Opfer2017}, Hamilton quaternion, coquaternion,
conectorine, and nectorine (refer to Table~\ref{table:1}) of which  the best known and most important is the Hamilton quaternion.\footnote{In
literature, the Hamilton's quaternion traditionally is referred to just as a quaternion. However, in the present paper by a `quaternion' we mean
one of the above mentioned quaternions.}  A new topic related to nonlinearities, namely, the roots of real  quaternions and  quaternionic
polynomials have been investigated by various methods \cite{Niven1942,Pogorui2004,deLeo2006,Ozdemir2009,Janovska2014,Janovska2017,Falcao2018}.
The fundamental theorem of algebra for quaternions formulated by I.~Niven~\cite{Niven1942} was used for this purpose in case of polynomials of
degree~$n$ as well as auxiliary (companion)  real  polynomial of degree $2n$ in root analysis \cite{Janovska2017}. It was shown that there may
be isolated zeros as well  as continuous zeros on spherical and hyperbolic surfaces, or no polynomial zeros at all. We have applied a different
approach to find square roots in 3D Clifford algebras~\cite{Dargys-Acus2020,Acus2026a}. Finally, quaternion algebra approach to nonlinear
Schr{\"o}dinger equation, in which a nonlinearity comes from square of quaternion modulus, was investigated in paper~\cite{Demontis2024}.

It is well-known that the square root of a general complex number  gives two complex  (plus/minus) numbers. However, an arbitrary rational power
of the complex number in general is a multi-valued quantity~\cite{Korn1961}. For example, cubic root of $-1$ gives three roots, $-1$ and two
complex conjugate roots $(-1\pm\ii\sqrt{3})/2$. Here we are interested in complexified/complex\footnote{
  In the paper we will not consider subtle difference between complex and complexified algebras~\cite{Marchuk2018} and will use the both terms interchangeably.} quaternions that are two times larger than  real
ones~\cite{Ward1997}. The main properties of biquaternions (complexified Hamilton quaternions) have been summarized
in~\cite{Sangwine2011,Berry2021}.  As far as we know, biquaternionic  square roots were considered just in a single paper~\cite{Sangwine2006} in
connection with $\sqrt{-1}$ that plays a key role in the Clifford-Fourier transform and Clifford
 algebra based wavelet theory \cite{Hitzer2013,Hitzer2022}. In Sec.~\ref{sec:3} we shall show  that
 complex quaternions (complexified  Hamilton quaternions, coquaternions, conectorines and nectorines) all are
 isomorphic to real 3-dimensional  Clifford algebra \cl{3}{0}. The roots of the latter have been analyzed
 in detail  in our work~\cite{Acus2026a}.  Therefore, we expect that similarly to roots in \cl{3}{0}
 the complexified quaternionic roots also are
 elaborate quantities and therefore have more than two $(\pm)$ square roots. In the present paper we have used
 the property that  both the real quaternions and the complex quaternions are isomorphic to real Clifford
 algebras.\footnote{Apart from mentioned non-commutative quaternionic algebras there are four additional commutative quaternionic
 algebras~\cite{Janovska2016}: tessarines, cotessarines, tangerines, and cotangerines. Since we are interested in
 relationship with the Clifford  algebra which is non-commutative, the named  commutative algebras are not considered in this paper.}
 Within this context the reader also may be  interested in quaternionic polynomial roots
 \cite{Opfer2017,Janovska2014,Janovska2017}.

In Sec.~\ref{sec:2}, main  properties of  real quaternions, the square roots of which are real numbers, are summarized, where the procedure to
extract the root from quaternion is explained too. In Sec.~\ref{sec:3}, the relations between  complex quaternions and  real Clifford algebra
\cl{3}{0} are presented that allow to find complex quaternion roots. In Sec.~\ref{sec:4}, examples of selected quaternionic roots calculated in
this way are inspected.

\section{Real quaternions and Clifford algebras}\label{sec:2}

\subsection{Kinds of real quaternions}\label{sec:2.1}

\begin{table}[t]
\caption{\footnotesize{Properties of real quaternionic elements and their  relation (isomorphism) with real Clifford algebras (CA), \cl{0}{2}
and \cl{2}{0}. Noncommutative products of basis elements can be obtained  from triple product, $\ii\jj\kk=\pm 1$, using multiplication either
  from left or right by $\ii$, $\jj$ and~$\kk$ in succession. Since isomorphism relation is transitive, the split quaternion (coquaternion),
  conectorine and nectorine represent mutually isomorphic algebras (the name indicates different choice of basis). }}
  \label{table:1}\vspace{2mm}
 $\begin{array}{llll}
 \textrm{Quaternion} &\textrm{Properties of quaternion basis}&\textrm{CA}&\mspace{-8mu}\textrm{Isomorphism} \\
\hline
\textrm{Hamilton}&\ii^2=\jj^2=\kk^2=-1,&\cl{0}{2}&\mspace{-8mu} \{1,\ii,\jj,\kk\}\cong  \\
\textrm{quaternion},\bbH&\ii\jj\kk=-1,\ii\jj=\kk,\jj\kk=\ii,\kk\ii=\jj & &\mspace{-8mu}\{1,\e{1},\e{2},\e{12}\}\\[5pt]
\textrm{Split quaternion}&\ii^2=-1,\jj^2=\kk^2=1,&\cl{2}{0}&\mspace{-8mu}\{1,\ii,\jj,\kk\}\cong \\
\textrm{(coquaternion)},\bbH_{\text{coq}}&\ii\jj\kk=-1,\ii\jj=-\kk,\jj\kk=\ii,\kk\ii=-\jj & &\mspace{-8mu}\{1,\e{12},\e{1},\e{2}\} \\[5pt]
\textrm{Conectorine},\bbH_{\text{con}}&\ii^2=\jj^2=1,\kk^2=-1,&\cl{2}{0}&\mspace{-8mu}\{1,\ii,\jj,\kk\}\cong\\
  &\ii\jj\kk=-1,\ii\jj=\kk,\jj\kk=-\ii,\kk\ii=-\jj& &\mspace{-10mu}\{1,\e{1},\e{2},\e{12}\} \\[5pt]
\textrm{Nectorine},\bbH_{\text{nec}}&\ii^2=1,\jj^2=-1,\kk^2=1, &\cl{2}{0}&\mspace{-8mu}\{1,\ii,\jj,\kk\}\cong  \\
 &\ii\jj\kk=1,\ii\jj=\kk,\jj\kk=\ii,\ii\kk=\jj & &\mspace{-10mu}\{1,\e{1},\e{12},\e{2}\}\\
\hline
\end{array}$
 \end{table}

 There are four kinds of real non-commutative quaternions~\cite{Opfer2017} as listed in Table~\ref{table:1}, which  also includes
 isomorphisms between quaternionic and Clifford algebra basis elements that will appear useful later.
 In the Table~\ref{table:multiplication},
respective multiplication tables of basis elements are presented.  The main difference comes from combinations of signs of squared quaternionic
basis vectors $\ii$, $\jj$, and $\kk$.  Also, the isomorphisms presented  in the Table~\ref{table:1}   may be  useful in quaternionic
computations by  Clifford algebra computer programs.\footnote{For real split quaternions
 we have used different isomorphic algebra, namely,  \cl{2}{0}, instead of \cl{1}{1} as proposed in~\cite{Opfer2017}. As we shall see  later,
 \cl{2}{0} allows  to represent all complexified quaternions by \cl{3}{0} algebra only.}  The Table~\ref{table:1}  shows that
 the quaternion~$q$ in both the Clifford algebra (CA) $\cl{p}{q}$ and the quaternion algebra (QA)
 may be represented similarly,
  \[\begin{split}
  &\text{CA}:\ Q=q_0+q_1\e{1}+q_2\e{2}+q_3\e{12},\\
   &\text{QA}:\ q=q_0+q_1\ii+q_2\jj+q_3\kk=q_0+\bq,
  \end{split}\]
where the coefficients are real numbers, $q_i\in\bbR$, not necessarily equal in different algebras.  $\e{1}$ and $\e{2}$ are orthonormal CA
basis vectors, and $\e{12}=\e{1}\w\e{2}$ is a bivector which represents a unit oriented plane constructed from outer product of orthonormal
vectors $\e{1}$ and $\e{2}$~\cite{Lounesto97}. In QA, frequently the quaternion is dealt with as a sum of scalar part, $q_0$, and vector part,
$\bq=q_1\ii+q_2\jj+q_3\kk$. Thus, the main difference between the two algebras comes from treatment of the vector space: In CA the vector space
is  two dimensional that contains the oriented plane $\e{12}$, while in QA the vector space is assumed to be  three dimensional, which consists
of three orthogonal basis vectors $\{\ii,\jj,\kk\}$. Table~\ref{table:1} shows the relations between two Clifford algebras, anti-Euclidean
\cl{0}{2} and Euclidean \cl{2}{0}, and $\{\ii,\jj,\kk\}$. The relations between $\ii$, $\jj$ and $\kk$ are fully expressed in multiplication
Table~\ref{table:multiplication}.

\begin{table}[t]
\caption{\footnotesize{Multiplication tables for quaternion algebras $\bbH$, $\bbH_{\textrm{coq}}$, $\bbH_{\textrm{con}}$ and
  $\bbH_{\textrm{nec}}$. The tables for $\bbH_{\textrm{coq}}, \bbH_{\textrm{con}}$ and $\bbH_{\textrm{nec}}$ are mutually
  isomorphic, while $\bbH$ is not isomorphic.}}
\label{table:multiplication}
\newcolumntype{R}{>{$}r<{$}}
\begin{tabular}{R|RRRR}\hspace{4.5mm}
\bbH  &1 &\ii &\jj &\kk\\ \hline
1  &1 &\ii &\jj &\kk\\
\ii &\ii &-1 &\kk &-\jj \\
\jj &\jj &-\kk &-1 &\ii \\
\kk &\kk &\jj &-\ii &-1 \\
\end{tabular}
\hspace{9.5mm}
\begin{tabular}{R|RRRR}
\bbH_{\textrm{coq}}  &1 &\ii &\jj &\kk\\ \hline
1  &1 &\ii &\jj &\kk\\
\ii &\ii &-1 &-\kk &-\jj \\
\jj &\jj &\kk &1 &\ii \\
\kk &\kk &-\jj &-\ii &1 \\
\end{tabular}
\newline\vspace{3mm}\\
\begin{tabular}{R|RRRR}
\bbH_{\textrm{con}}  &1 &\ii &\jj &\kk\\ \hline
1  &1 &\ii &\jj &\kk\\
\ii &\ii &1 &\kk &\jj \\
\jj &\jj &-\kk &1 &-\ii \\
\kk &\kk &-\jj &\ii &-1 \\
\end{tabular}
\hspace{12mm}
\begin{tabular}{R|RRRR}
\bbH_{\textrm{nec}}  &1 &\ii &\jj &\kk\\ \hline
1  &1 &\ii &\jj &\kk\\
\ii &\ii &1 &-\kk &\jj \\
\jj &\jj &-\kk &-1 &\ii \\
\kk &\kk &-\jj &-\ii &1 \\
\end{tabular}
\end{table}

The quaternionic basis multiplication rules in Table~\ref{table:multiplication} permit to introduce a product of two quaternions, $q=q_0+\bq$
and $p=p_0+\bp$, in following form,
\begin{equation*}
qp\equiv(q_0,\bq)(p_0,\bp)=(q_0p_0-\bq.\bp,q_0\bp+p_0\bq+\bq\times\bp),
\end{equation*}
where $\bq.\bp=q_1p_1\ii^2+q_2p_2\jj^2+q_3p_3\kk^2$ is the scalar product and $\bq\times\bp$ is the vector product. The latter usually is
expressed in a form of determinant,
\begin{equation*}
\bq\times\bp=
\begin{vmatrix}
\ii&\jj&\kk\\
q_1&q_2&q_3\\
p_1&p_2&p_3\\
\end{vmatrix}
=(q_2p_3-q_3p_2)\ii+(q_3p_1-q_1p_3)\jj+(q_1p_2-q_2p_1)\kk.
\end{equation*}
If quaternion product $qp$ is recopied in Clifford algebra notation in accord to the last column  of the Table~\ref{table:1},  $q\to
Q=q_0+q_1\e{1}+q_2\e{2}+q_3\e{12}$ and $p\to P=q_0+p_1\e{1}+p_2\e{2}+p_3\e{12}$,  then one would notice that quaternionic product is equivalent
to multivector geometric product $QP$ for all quaternions listed in the Table~\ref{table:1}. The reader also should notice that after
replacement of $\{\ii,\jj,\kk\}$ by respective vectors, $\e{1}$ and $\e{2}$, and bivector $\e{12}$ from Table~\ref{table:1} will result in
different multivectors Q and P, and therefore in different square roots for all four kinds of quaternions as shown in Subsec~\ref{sec:4.1}. From
all this follows that the structure of two dimensional Clifford algebras, in general, is richer than that of quaternions, thus, in order to
represent all four quaternions it suffice  to make use of two Clifford algebras, \cl{0}{2} and \cl{2}{0}, only. In physics, the triad
$\{\ii,\jj,\kk\}$ frequently is interpreted as an orthogonal vector basis in three dimensional physical space. Then, the following relation
between $\bq\times\bp$ and outer product $\bq\w\bp$ in the Euclidean algebra \cl{3}{0} can be established: $\bq\times\bp=-I_3\bq\w\bp$, where
$I_3=\e{123}\in\cl{3}{0}$ is the pseudoscalar~\cite{Doran03}. From this relation follows that $\kk=\ii\times\jj\equiv\ii\jj=I_3\e{21}$ for
$\bbH$ as well as similar relations after cyclic permutations, $\jj=I_3\e{13}$ and $\ii=I_3\e{32}$.

 The conjugate quaternion $q^{*}$ is defined by $q^{*}=q_0-\bq=q_0-q_1\ii-q_2\jj-q_3\kk$. Then, the product
 $q^{*}q=qq^{*}=q_0^2-\bq_0^2=q_0^2-q_1^2\ii^2-q_2^2\jj^2-q_3^2\kk^2$ may be positive, negative or zero.
The norm  is defined by $|q|=\sqrt{qq^{*}}$ if $qq^{*}>0$. For Hamilton quaternions, the norm squared is always positive,
$q^{*}q=q_0^2+q_1^2+q_2^2+q_3^2>0$. The inverse of quaternion is defined by $q^{-1}=q^{*}/(q^{*}q)$ which always exists in a case of Hamilton
quaternion. For remaining  quaternions  the norm exists only in a domain where $qq^{*}>0$.

\subsection{Multivector square roots in  two dimensional algebras which are related to real quaternions}\label{sec:2.2}

 A square root $\m{A}=\pm\sqrt{\m{B}}$ of general multivector $\m{B}=b_0+b_1 \e{1}+ b_2 \e{2} + b_3 \e{12}$ for algebras  $\cl{2}{0}$, $\cl{1}{1}$ and $\cl{0}{2}$ has the form $\m{A}=s+v_1 \e{1} + v_2
\e{2} + S \e{12}$. A simple analysis shows that coefficients $(s,S)$ are given by formulas and conditions,
\begin{align*}
\mkern-36mu&\begin{cases}
  \Bigl(s=\pm\frac{1}{\sqrt{2}}\sqrt{b_0-\sqrt{D}},S=\pm\frac{1}{\sqrt{2}} \frac{b_{3}}{\sqrt{b_0-\sqrt{D}}}\Bigr),\quad
  \textrm{if}\quad b_0-\sqrt{D} > 0 \ \textrm{and}\ D\ge 0, \\
\Bigl(s=\pm\frac{1}{\sqrt{2}}\sqrt{b_0+\sqrt{D}},S=\pm\frac{1}{\sqrt{2}} \frac{b_{3}}{\sqrt{b_0+\sqrt{D}}}\Bigr),\quad \textrm{if}\quad
b_0+\sqrt{D} > 0 \ \textrm{and}\ D\ge 0,
\end{cases}
\end{align*}
where the determinant of $\m{B}$ is defined by
\begin{align*}
& D= \begin{cases}
     b_0^2-b_1^2-b_2^2+b_3^2, \quad \textrm{for}\ \cl{2}{0},\\
     b_0^2-b_1^2+b_2^2-b_3^2, \quad \textrm{for}\ \cl{1}{1},\\
     b_0^2+b_1^2+b_2^2+b_3^2, \quad \textrm{for}\ \cl{0}{2}.
   \end{cases}
\end{align*}

{\it The case $s\neq 0$}. The coefficients $v_1,v_2$ in $\m{A}$ then are given by formulas
\begin{align*}
&v_1=\frac{b_1}{2 s},\quad v_2=\frac{b_2}{2 s}.
\end{align*}

{\it The case $s=0$}. When $b_0-\sqrt{D} = 0$ or $b_0+\sqrt{D} = 0$ and $b_1=b_2=b_3=0$ the coefficients $v_1, v_2$ and $S$ are related by
single equation of a type $\pm v_1^2\pm v_2^2\pm b_0\pm S^2=0$, which can be  solved for any of coefficients of $v_1, v_2$ or $S$, the remaining
two assuming as free parameters. For example, if we solve for $S$, then the coefficients $v_1$ and $v_2$  may be considered as free parameters
and the square root for each of algebra can be written as
\begin{align*}
\mkern-36mu& \m{A}=\begin{cases}
v_1 \e{1} + v_2 \e{2} \pm\sqrt{-b_0+v_1^2+v_2^2} \e{12},&\   \textrm{for}\ \cl{2}{0},  \quad \textrm{if}\quad b_1=b_2=b_3=0,\\
v_1 \e{1} + v_2 \e{2} \pm\sqrt{b_0-v_1^2+v_2^2} \e{12},&\   \textrm{for}\ \cl{1}{1}, \quad \textrm{if}\quad b_1=b_2=b_3=0,\\
v_1 \e{1} + v_2 \e{2} \pm\sqrt{-b_0-v_1^2-v_2^2} \e{12},&\   \textrm{for}\ \cl{0}{2}, \quad \textrm{if}\quad b_1=b_2=b_3=0.
\end{cases}
\end{align*}
Since the coefficient $S$ must be real, i.e. the root exists only when expressions under square root are positive.

\vspace{3mm} Example in case $s\ne 0$. The square roots  of $\m{B}=6+2 \e{1}+3\e{2}-4\e{12}$  are
\begin{align*}
\mkern-36mu& \m{A}=\begin{cases}
\pm\frac{1}{\sqrt{2 (6+\sqrt{39})}}(6+\sqrt{39}+2 \e{1}+3 \e{2}-4 \e{12})&  \cl{2}{0},\\
  \pm\frac{1}{\sqrt{2}} (1+2 \e{1}+3 \e{2}-4 \e{12}) \textrm{ and }
  \pm\frac{1}{\sqrt{22}}(11+2 \e{1}+3 \e{2}-4 \e{12})& \cl{1}{1}, \\
  \pm\frac{1}{\sqrt{2 (6+\sqrt{65})}} (6+\sqrt{65}+2 \e{1}+3 \e{2}-4 \e{12})& \ \cl{0}{2}.
\end{cases}
\end{align*}
Note that in $\cl{1}{1}$ there are four roots, because the both conditions, $b_0+\sqrt{D}=11>0$ and $b_0-\sqrt{D}=1>0$, are satisfied.
 Examples of square roots for real quaternions are presented in Subsec.~\ref{sec:4.1}.
 In the next section, complex quaternions, i.e. quaternions with complex coefficients, are considered.

\section{Complex  quaternions and isomorphism with \cl{3}{0}}\label{sec:3}
Main properties of complex  quaternions are surveyed,  for this purpose employing  the  Hamilton quaternion $\bbH$. The isomorphisms between
various complex quaternions and real 3D Clifford algebra \cl{3}{0} is discussed.  More can be found in~\cite{Ward1997,Altmann2005}.

It is well-known that all complex Clifford algebras which belong to  same vector space dimension are isomorphic~\cite{Lounesto97}. Below we show
that complex Clifford algebras of vector space dimension $n=2$ are isomorphic to a single real $n=3$ Clifford algebra. Indeed, since real
Clifford algebras $\cl{2}{0}$ and $\cl{1}{1}$ are isomorphic, it is sufficient to investigate the complexificaton of $\cl{2}{0}$ and $\cl{0}{2}$
only. Remembering that  complex numbers are represented by $\cl{0}{1}$, the complexification  of  even  $n=2$ algebras, $\cl{2}{0}$ and
$\cl{0}{2}$, is equivalent to tensor product of real algebras $\cl{0}{1}\otimes\cl{2}{0}\cong\cl{3}{0}$ and
$\cl{0}{1}\otimes\cl{0}{2}\cong\cl{1}{2}$, respectively (why similar consideration can't be applied to odd $n$ see explanation in the next
footnote). Since real Clifford algebras $\cl{1}{2}$ and $\cl{3}{0}$ are isomorphic, it follows that all complexified $n=2$ algebras (and,
consequentially, all named complexified quaternions) may be assumed to be isomorphic to $\cl{3}{0}$.

\subsection{Complexified  Hamilton quaternion}\label{sec:3.1}
The complexified Hamilton quaternion  is denoted by $\bbH^{\bbC}$ and is referred to a biquaternion as well. The real Hamilton quaternion $\bbH$
is isomorphic to \cl{0}{2}, correspondingly, the biquaternion $\bbH^{\bbC}$ is isomorphic to  complexified CA $\textit{Cl}_2^\bbC$. The real,
$q$, and complexified, $Q$, Hamilton quaternions  are defined similarly,
\[\begin{split}
q&=q_0+q_1\ii+q_2\jj+q_3\kk,\qquad\ii^2=\jj^2=\kk^2=-1,\quad q_l\in\bbR,\quad\textit{real quaternion $\bbH$} \\
Q&=q_0+q_1\ii+q_2\jj+q_3\kk,\qquad\ii^2=\jj^2=\kk^2=-1,\quad q_l\in\bbC,\quad\textit{biquaternion $\bbH^{\bbC}$}\\
\end{split}\] except that coefficients of the biquaternion (complexified Hamilton quaternion)  are complex numbers:
 \[q_0=q_{0r}+Iq_{0i},\quad q_1=q_{1r}+Iq_{1i},\quad q_2=q_{2r}+Iq_{2i},\quad q_3=q_{3r}+Iq_{3i},\]
where $q_{li}$  and $\ q_{lr}\in\bbR$ and $I\equiv\sqrt{-1}$ stands for a standard imaginary unit, $I^2=-1$, since  $\ii$ is already reserved
for quaternionic basis $\{\ii,\jj,\kk\}$. It is assumed that the (bi)quaternionic elements  $\ii$, $\jj$ and $\kk$ commute with the imaginary
unit~$I$. Now, one has two kinds of conjugations: complex and quaternionic. Under complex conjugation the coefficient, for example, of
$q_0=q_{0r}+Iq_{0i}$ transforms to $q_0^*=q_{0r}-Iq_{0i}$. The quaternionic conjugation, which will be denoted by a five-pointed star $\star$,
changes signs  of $\ii$, $\jj$ and $\kk$ only. Thus, a complexified quaternion $Q$ in a full form can be written  as
\[\begin{split}
Q=\;&(q_{0r}+Iq_{0i})+(q_{1r}+Iq_{1i})\ii+(q_{2r}+Iq_{2i})\jj+(q_{3r}+Iq_{3i})\kk\equiv \\
&\qquad\qquad\qquad q_0+q_1\ii+q_2\jj+q_3\kk,\\
Q^*=&(q_{0r}-Iq_{0i})+(q_{1r}-Iq_{1i})\ii+(q_{2r}-Iq_{2i})\jj+(q_{3r}-Iq_{3i})\kk,\quad\\
Q^\star=&(q_{0r}+Iq_{0i})-(q_{1r}+Iq_{1i})\ii-(q_{2r}+Iq_{2i})\jj-(q_{3r}+Iq_{3i})\kk.\\
\end{split}\]
The first expression in $Q$  shows that the complexified quaternion can be written as a sum of two  Hamilton  quaternions, $ Q=q_r+q_i I$ where
$q_r=q_{0r}+q_{1r}\ii+q_{2r}\jj+q_{3r}\kk$ and similarly $q_i=q_{0i}+q_{1i}\ii+q_{2i}\jj+q_{3i}\kk$ are real quaternions (hence the name
`biquaternion'). Contrary to real quaternion, the product of biquaternion and conjugate biquaternion is a complex number
\[QQ^\star=Q^\star Q=-\sum_{p=0}^{p=3}(q_{pi}-Iq_{pr})^2. \]
If imaginary parts are equated to zero, $q_{pi}=0$, the formula returns quaternion semi-norm: $QQ^\star=-(q_{0r}^2+q_{1r}^2+q_{2r}^2+q_{3r}^2)$.
The norm squared  of a full biquaternion can be defined by
\[|Q|^2=\tfrac12\big(Q (Q^{\star})^{*}+Q^{*}Q^\star\big)=(q_{0r}^2+q_{0i}^2)+(q_{1r}^2+q_{1i}^2)+(q_{2r}^2+q_{2i}^2)+(q_{3r}^2+q_{3i}^2), \]
which is positive. If $P$ and $Q$ are the biquaternions then $(PQ)^{*}=Q^{*}P^{*}$, $(PQ)^\star=P^{\star}Q^\star$,  and
$(Q^\star)^{*}=(Q^*)^{\star}$.

Matrix form of biquaternionic basis is represented by following complex matrices,
\[\hat{1}=\begin{pmatrix}1&0\\0&1\end{pmatrix},\quad
 \hat{\ii}=\begin{pmatrix}I &0\\0&-I\end{pmatrix},\quad
 \hat{\jj}=\begin{pmatrix}0&1\\-1&0\end{pmatrix},\quad
 \hat{\kk}=\begin{pmatrix}0&I\\ I&0\end{pmatrix},\quad I\equiv\sqrt{-1}\]
squares of which give  unit matrices,  $\{\hat{1},-\hat{1},-\hat{1},-\hat{1}\}$, as indicated by hats. From these matrices,  a general
biquarternion in complex (now specified by $I$) matrix representation follows,
\[\hat{Q}=q_0\hat{1}+q_1\hat{\ii}+q_2\hat{\jj}+q_3\hat{\kk}=
 \begin{pmatrix}q_0+I q_1&q_2+I q_3\\-q_2+I q_3&q_0-I q_1\end{pmatrix},\quad(q_0,q_1,q_2,q_3)\in\bbC. \]
Thus, all in all, there are 8 independent elements in a general complex  quaternion. As we shall see this set is isomorphic to real \cl{3}{0}
algebra, $\bbH^\bbC\cong\cl{3}{0}$. This can be verified by direct comparison of multiplication tables of both, $\bbH^\bbC$ and $\cl{3}{0}$,
algebras (also see the next subsection\ref{sec:3.2}).

Multiplication of complex quaternions is done similarly as in case of complexified multivectors, i.e., it is  assumed that  $I$ and
$\{\ii,\jj,\kk\}$  commute and $I^2=-1$, while mutually anticommuting products between $\ii$, $\jj$ and $\kk$ are to be  calculated  according
to Table~\ref{sec:2}. For example,

 1) $\big(1+\ii+(1-I2)\jj\big)(3-I\jj)=(5+I)+3\ii+(3-7 I)\jj-I\kk$;\\
\hspace*{16mm}  2) $\big((\ii+\jj+\kk)+I(\jj-\kk)\big)^2=-1$; \ \ \ 3) $(1+I\ii)(1-I\ii)=0$.\\
The first example shows that, in general, the product of two biquaternions is a biquaternion too. The second example is equivalent to square
root of $-1$ in $\bbH^{\bbC}$, i.e., $\sqrt{-1}=(\ii+\jj+\kk)+I(\jj-\kk)$. This can be checked by squaring the left and right sides and using
just  described multiplication properties for complex quaternions. The third example shows that, contrary to real Hamilton quaternions, some
biquarternions may be zero divisors~\cite{Ward1997}. Also, the complexified quaternions may be nilpotent, for example, $(\ii+I\jj)^2=0$.

Below we shall show that there exists isomorphisms between  Euclidean Clifford algebra \cl{3}{0} and individual complexified quaternionic
algebras.


\subsection{\label{sec:3.2}Isomorphisms between $\cl{3}{0}$ and complex quaternions}

The real quaternion basis, $\{1,\ii,\jj,\kk\}$, consists of four elements. In complexified quaternion algebras the number of basis elements
doubles.  We will use the isomorphism between complex quaternions and real Clifford algebra $\cl{3}{0}$, where the additional basis elements are
products of $\{1,\ii,\jj,\kk\}$ and $I$ what, all in all, yields eight basis elements $\{1,\ii,\jj,\kk,I,I\ii,I\jj,I\kk\}$ that are isomorphic
to basis vectors of real \cl{3}{0} algebra.\footnote{{A general complexified Clifford algebra  $\cl{p}{q}$ of vector space dimension $n=p+q$ is
isomorphic to a  larger  real  Clifford algebra of  dimension $n+1$ only if $n$ is even.
 For odd $n$, the complexified algebra is isomorphic to the so-called} extended Clifford
 algebra~\cite{Marchuk2018}. For example, if we take real $\cl{3}{0}$ algebra, which can be understood
 as a tensor product $\cl{0}{1}\otimes\cl{2}{0}$, then after complexification we write
 $\cl{0}{1}\otimes\cl{3}{0}$ which is equivalent to  $\cl{0}{1}\otimes\cl{0}{1}\otimes\cl{2}{0}$.
 It is easy to see  that this triple tensor product can't be isomorphic to any  Clifford
 algebra.  Indeed, in equivalent tensor product the pairs of basis vectors that come from tensor product   $\cl{0}{1}\otimes\cl{0}{1}$
 of one dimensional algebras  mutually commute. On the other hand, the definition of Clifford algebra assumes
anti-commuting vector basis. Therefore, the tensor product $\cl{0}{1}\otimes\cl{3}{0}$ is isomorphic
 to some extended real Clifford algebra.}
To find out the isomorphism between algebra generators, in \cl{3}{0} multiplication table at first
 the isomorphism was determined  in an upper $4\times 4$ diagonal block  related  to
  $\{1,\ii,\jj,\kk\}$ and then in a lower $4\times 4$ diagonal block  related  to  $\{I,I\ii,I\jj,I\kk\}$.
 Remaining non-diagonal $4\times 4$ blocks then tune in to correct values automatically.
 In Tables \ref{table:3}-\ref{table:6}, the isomorphisms  between complex quaternions and \cl{3}{0} are given. The first line indicates sign of
 squares of the generators in the second and third line.
\vspace{4mm}

$\blacktriangledown$ The isomorphism, $\bbH^{\bbC}\cong\cl{3}{0}$, between complexified Hamilton quaternion $\bbH^{\bbC}$ and real Clifford
algebra \cl{3}{0}.
\begin{table}[h]
\caption{\footnotesize{Complex Hamilton quaternion \label{table:3}}}
\begin{center}
\begin{tabular}{l||l |l |l |l ||l |l |l |l}
$(\sharp)^2$ &$+$&$-$&$-$&$-$&$-$&$+$&$+$&$+$\\ \hline
 $\bbH^{\bbC}$ &1 &$\ii$   &$\jj$     &$\kk$   &$I$ &$I\ii$   &$I\jj$     &$I\kk$ \\ \hline
 \cl{3}{0} &1 &$\e{12}$&$-\e{13}$ &$\e{23}$  &$\e{123}$ &$-\e{3}$&$-\e{2}$ &$-\e{1}$
\end{tabular}
\end{center}
\end{table}
 Both the isomorphism in Table~\ref{table:3}{}  and comprehensive  analysis how to extract the square root from  multivector~\cite{Acus2026a} in \cl{3}{0}
 (also  see Sec.~\ref{sec:4.6} and Algorithm~\ref{AlgForSqrt30} in the Appendix), allow  to find
 all square roots of a complex quaternion $\bbH^{\bbC}$ immediately. For example, in
 \cl{3}{0} one finds out that $\e{1}$ has four roots, namely, $\sqrt{\e{1}}=\pm\tfrac12(1+\e{1}+\e{23}-I_3)$ and
 $\sqrt{\e{1}}=\pm\tfrac12(1+\e{1}-\e{23}+I_3)$, where $I_3=\e{123}$ is the pseudoscalar in \cl{3}{0}. With the help of the isomorphism
 in Table~\ref{table:3} it is easy to transform the roots to $\bbH^{\bbC}$ algebra,
\[\sqrt{-I\kk}=\begin{cases}
&\pm\tfrac12(1-I)(1+\kk),\\
&\pm\tfrac12(1+I)(1-\kk).
\end{cases}\]
More examples are presented in Sec.~\ref{sec:4}. The remaining quaternions (coquaternion,  conectorine, and nectorine)  can be complexified in a
similar way too.
 \vspace{4mm}

 $\blacktriangledown$ Isomorphism between complexified coquaternion,  $\bbH^{\bbC}_{\text{coq}}$,
 and real Clifford algebra \cl{3}{0}.
 \begin{table}[h!]
\caption{\footnotesize{Complexified  coquaternion \label{table:4}}}
\begin{center}
\begin{tabular}{l||l |l |l |l ||l |l |l |l}
$(\sharp)^2$  &$+$&$-$&$+$&$+$&$-$&$+$&$-$&$-$\\ \hline
 $\bbH^{\bbC}_{\text{coq}}$ &1 &$\ii$   &$\jj$     &$\kk$   &$I$ &$I\ii$   &$I\jj$     &$I\kk$ \\ \hline
 \cl{3}{0} &1 &$\e{23}$&$\e{2}$ &$\e{3}$  &$\e{123}$ &$-\e{1}$&$-\e{13}$ &$\e{12}$
\end{tabular}
\end{center}
\end{table}
\vspace{4mm}

\newpage
 $\blacktriangledown$ Isomorphism between  complexified conectorine, $\bbH^{\bbC}_{\text{con}}$,
 and real Clifford algebra \cl{3}{0}.
 \begin{table}[h]
\caption{\footnotesize{Complexified  conectorine \label{table:5}}}
\begin{center}
\begin{tabular}{l||l |l |l |l ||l |l |l |l}
$(\sharp)^2$  &$+$&$+$&$+$&$-$&$-$&$-$&$-$&$+$\\ \hline
 $\bbH^{\bbC}_{\text{con}}$ &1 &$\ii$   &$\jj$     &$\kk$   &$I$ &$I\ii$   &$I\jj$     &$I\kk$ \\ \hline
 \cl{3}{0} &1 &$\e{1}$&$\e{2}$ &$\e{12}$  &$\e{123}$ &$\e{23}$&$-\e{13}$ &$-\e{3}$
\end{tabular}
\end{center}
\end{table}
\vspace{4mm}

 $\blacktriangledown$ Isomorphism between complexified nectorine, $\bbH^{\bbC}_{\text{nec}}$,
 and real Clifford algebra \cl{3}{0}.
\begin{table}[h]
\caption{\footnotesize{Complexified  nectorine \label{table:6}}}
\begin{center}
\begin{tabular}{l||l |l |l |l ||l |l |l |l}
$(\sharp)^2$  &$+$&$+$&$-$&$+$&$-$&$-$&$+$&$-$\\ \hline
 $\bbH^{\bbC}_{\text{nec}}$ &1 &$\ii$   &$\jj$     &$\kk$   &$I$ &$I\ii$   &$I\jj$     &$I\kk$ \\ \hline
 \cl{3}{0} &1 &$\e{1}$&$\e{12}$ &$\e{2}$  &$\e{123}$ &$\e{23}$&$-\e{3}$ &$-\e{13}$
\end{tabular}
\end{center}
\end{table}

The Tables \ref{table:3}-\ref{table:6} show that all four complexifed quaternion algebras are isomorphic to  \cl{3}{0}. Properties of the square
roots of multivectors in the real Euclidean algebra \cl{3}{0}  have been analyzed in detail in paper~\cite{Acus2026a}, where we have shown that
there may be isolated and continuous square roots that may contain  real parameters, or for some multivectors the roots may be absent at all.
Examples are presented in Sec.~\ref{sec:4}.

 Conclusion: The method used  in the paper is based on isomorphism between  non-commutative real  and complexified quaternions
 (Hamilton quaternion, coquaternion, conectorine and nectorine) and  real Clifford algebras  \cl{0}{2}, \cl{2}{0} and
 \cl{3}{0}, respectively. Square root properties of the latter have been analyzed in detail in \cite{Acus2026a}.
  The isomorphisms between quaternions and
 multivectors presented in the tables may be useful in transforming quaternionic equations  to equivalent ones in the Clifford algebra, instead of deducing
 the equations anew.

\section{Appendix: Examples of square roots for real and complex quaternions}\label{sec:4}
To extract the square root from real or complex quaternion, at first, the quaternion  should be transformed to respective multivector (see
tables in Subsec.~\ref{sec:2.2} and \ref{sec:3.2}), and then the root algorithm applied to the multivector. Subsequently, transformation of the
obtained roots  from Clifford algebra back to respective quaternionic form will give the final answer. Below  a number of example of
quaternionic square roots obtained in this way  is presented for all  kinds of real and  complex quaternions.

\subsection{Square roots of real quaternions}\label{sec:4.1}

$\blacktriangledown$ Let the  Hamilton quaternion, coquaternion, conectorine and nectorine be of the same simple shape, that is $q=1+\kk$.

Roots of  Hamilton quaternion. The transformation to CA (Table~\ref{table:1}) gives  multivector $\m{B}=1+\e{12}$ with the coefficients
$b_0=b_3=1$ and $b_1=b_3=0$. Since the determinant $D=2>0$, we find (Subsection~\ref{sec:2.2})
 \[\big(s=\tfrac{1}{\sqrt{2}}\sqrt{1+\sqrt{2}},\ S=\tfrac{1}{\sqrt{2}} \tfrac{1}{\sqrt{1+\sqrt{2}}}\big)\quad\text{and}\quad v_1=v_2=0.\]
 Then the square root of multivector $\m{B}$ is
 \[\m{A}=\sqrt{\m{B}}= \pm(s+v_1\e{1}+S\e{12})=\pm\frac{1}{\sqrt{2}\sqrt{1+\sqrt{2}}}\big(1+\sqrt{2}+\e{12}\big).\]
After transformation back (Table~\ref{table:1}), we have the root of Hamilton quaternion,
\[\bbH:\sqrt{q}=\pm\frac{1}{\sqrt{2}\sqrt{1+\sqrt{2}}}\big(1+\sqrt{2}+\kk\big).\]
In a similar way we can find the roots of the remaining quaternions,
\[\begin{split}
&\bbH_{\textrm{coq}}: \sqrt{q}=\pm(1+\kk)/\sqrt{2},\\
&\bbH_{\textrm{con}}:\sqrt{q}=\pm\frac{1}{\sqrt{2}\sqrt{1+\sqrt{2}}}\big(1+\sqrt{2}+\kk\big),\\
&\bbH_{\textrm{nec}}:\sqrt{q}= \pm(1+\kk)/\sqrt{2}.\\
\end{split}\]
After squaring of the roots one obtains the initial multivector, $q=1+\kk$. In general, all roots may be different, or there be no roots at all.
\vspace{2mm}

$\blacktriangledown$ Roots of real quaternion $\m{B}=1+2\ii+3\jj+4\kk$ are,
\[\begin{split}
&\bbH:\quad\sqrt{q}=\pm\tfrac{\sqrt{2}}{2\sqrt{1+\sqrt{30}}}\big(1+\sqrt{30}+2\ii+3\jj+4\kk \big); D=30>0,\\
&\bbH_{\textrm{coq}}: \text{no roots because the determinant}\ D=-20<0,\\
&\bbH_{\textrm{con}}:\sqrt{q}=\pm\tfrac{1}{\sqrt{6}} (3+2\ii+3\jj+4\kk); D=4>0,\\
&\bbH_{\textrm{nec}}: \text{no roots because the determinant}\ D=-20<0.\\
\end{split}\]

\subsection{Square roots of a complexified Hamilton quaternion $\bbH^{\bbC}$}\label{sec:4.2}
In \cl{3}{0}, the roots of basis vector $\e{1}$ are $\m{A}_{1,2}=\sqrt{\e{1}}=\pm\tfrac12(1+\e{1}+\e{23}-I_3)$ and
$\m{A}_{3,4}=\sqrt{\e{1}}=\pm\tfrac12(1+\e{1}-\e{23}+I_3)$, $I_3=\e{123}$ \cite{Acus2026a}. From  Table~\ref{table:3} it follows that
$\sqrt{\e{1}}\to\sqrt{-I\kk}$. Then, after transformation of right hand sides of $\m{A}_{1,2}$ and $\m{A}_{3,4}$ to complex  quaternions  we
obtain the roots of complex quaternion $-I\kk$,
\[\sqrt{-I\kk}=\begin{cases}
&\pm\tfrac12(1-I\kk+\kk-I),\\
&\pm\tfrac12(1-I\kk-\kk+I).
\end{cases}\]
The squares of right-hand  side expressions give $-I\kk$.
 \vspace{2mm}

More examples.\\
 \textit{Example 1}. Complex quaternion $\m{B}=-(2+I)\kk\in\bbH^{\bbC}$ has four roots,
\[\begin{split}
 & \m{A}_{1,2}=\pm\tfrac12\sqrt{2+\sqrt{5}}\;\big(-2+\sqrt{5}-I\kk+(-2+\sqrt{5})\kk-I\big),\\
 & \m{A}_{3,4}=\pm\tfrac12\sqrt{-2+\sqrt{5}}\;\big(2+\sqrt{5}-I\kk-(2+\sqrt{5})\kk+I\big). \\
 \end{split}\]
 \textit{Example 2}.  $\m{B}=(-1+\tfrac{1}{2}I)-(1+I)\ii$ has four roots too,
\[\m{A}_{1,2}=\pm\tfrac12(-1-I+2\ii),\quad
 \m{A}_{3,4}=\pm\tfrac12\big(-2 I+(1-I)\ii\big).\\
 \]

 \noindent\textit{Example 3}.  $\m{B}=-1+I$ has two roots,
 \[ \m{A}_{1,2}=\pm\Big(\sqrt{-\tfrac{1}{2}+\tfrac{1}{\sqrt{2}}}+I\sqrt{\tfrac{1}{2}+\tfrac{1}{\sqrt{2}}}\Big).\]

\noindent\textit{Example 4}. The roots of $\m{B}=1-(\ii-\jj)I -\kk$ are
 \[\m{A}_{1,2}=\pm(I+\ii+\jj-I\kk)/\sqrt{2},\quad
  \m{A}_{3,4}=\pm\big(3- I\ii+I\jj-\kk)/\sqrt{6}\big).
\]

 \noindent\textit{Example 5}. However, $\m{B}=\ii-I\kk$ has no roots.

\subsection{Square roots of complex coquaternion, $\bbH^{\bbC}_{\text{coq}}$}\label{sec:4.3}

 \textit{Example 1}.  Complexified coquaternion $\m{B}=-(2+I)\ii$ has four roots,
\[\begin{split}
 & \m{A}_{1,2}=\pm\tfrac12\Big(\sqrt{-2+\sqrt{5}\;} -I\sqrt{2+\sqrt{5}\;}  \Big)(1+\ii),\\
 & \m{A}_{3,4}=\pm\tfrac12\Big(\sqrt{-2+\sqrt{5}\;} +I\sqrt{-2+\sqrt{5}\;} \Big)(-1+\ii).\\
 \end{split}\] The squares of roots give initial  coquaternion $\m{B}$.


\subsection{Square roots of complex conectorine, $\bbH^{\bbC}_{\text{con}}$}\label{sec:4.4}

\noindent\textit{Example 1}. Conectorine $\m{B}=(-1+I/2)+(1+I)\kk$ has four roots,
\[\begin{split}
 &\m{A}_{1,2}=\pm\tfrac12(-1-I+2\kk),\\
 &\m{A}_{3,4}=\pm\tfrac12\big(-2I+(1-I)\kk\big).
 \end{split}\]

\noindent\textit{Example 2}. Conectorine $\m{B}=(1+I\ii+I\jj+\kk)$ has four roots too,
\[\begin{split}
 &\m{A}_{1,2}=\pm(I+\ii+\jj-I\kk)/\sqrt{2}\;,\\
 &\m{A}_{3,4}=\pm(3+I\ii+I\jj+\kk)/\sqrt{6}\;.
 \end{split}\]

\subsection{Square roots of complex nectorine, $\bbH^{\bbC}_{\text{neq}}$}\label{sec:4.5}

\noindent\textit{Example 1}. Nectorine $\m{B}=(1-2I)\ii$ has two roots,
\[\m{A}_{12}=\pm\tfrac12\sqrt{2+\sqrt{5}}\Big(-2+\sqrt{5}-I+\big(1+(-2+\sqrt{5})I\big)\ii\Big).\]

\noindent\textit{Example 2}. Nectorine $\m{B}=1+I\ii+\jj+I\kk$ has four roots,
\[\begin{split}
&\m{A}_{1,2}=\pm(I+\ii-I\jj+\kk)/\sqrt{2}\;,\\
&\m{A}_{3,4}=\pm(3+I\ii+\jj+I\kk)/\sqrt{6}\;.
 \end{split}\]

\subsection{Square root algorithm for real $\cl{3}{0}$}\label{sec:4.6}

Algorithm~\ref{AlgForSqrt30}  finds  all square roots (both discrete and continuum) of real multivector~$\m{B}$ in \cl{3}{0}.
 It can be applied to real multivectors in numerical and symbolic forms.
  For more details,  the reader is directed to paper~\cite{Acus2026a}.
 In the Algorithm~\ref{AlgForSqrt30}, $\m{B}$  is input and  $\m{A}=s+\mathbf{v}+(S+\mathbf{V})\e{123}$ is output
 multivector,  where  $s$ and $S$ are scalars,
 $\mathbf{v}=v_1\e{1}+v_2\e{2}+v_3\e{3}$ and $\mathbf{V}=V_1\e{1}+V_2\e{2}+V_3\e{3}$ are vectors. If output coefficients
 do not receive concrete values then  they are considered to be free parameters. Input multivector coefficients
 satisfy $b_0=\mathbf{v^2-V^2}$ and $b_{123}=2(\mathbf{v\cdot V})$, that have following geometrical  interpretation:
 The first expression represents a pair of concentric spheres, where the input scalar coefficient  $b_0$ determines
 difference between sphere radii, $\mathbf{|v|}$ and $\mathbf{|V|}$. The second expression controls an angle
  between the vectors $\mathbf{v}$ and $\mathbf{V}$.

\begin{algorithm}[H]
\DontPrintSemicolon
\SetAlgoLined
\SetNoFillComment \LinesNotNumbered
\SetKwBlock{Begin}{Begin}{}
\SetKwInput{KwInput}{Input} \SetKwInput{KwOutput}{Output} \SetKwProg{Sqrt}{Sqrt}{}{} \Sqrt{$(\m{B})$}{
    \KwInput{$\m{B}=b_0+b_1\e{1}+b_2\e{2}+b_3\e{3}+b_{12}\e{12}+b_{13}\e{13}+b_{23}\e{23}+b_{123}\e{123}$}
    \KwOutput{$\m{A}_{i_k}=s_i+v_{i_1}\e{1}+v_{i_2}\e{2}+v_{i_3}\e{3}+
  (S_i+V_{i_1}\e{1}+V_{i_2}\e{2}+V_{i_3}\e{3})\e{123}$}
 \vskip 2pt
  {\scriptsize \tcc{Initialization}}
$b_{S}=b_{0}^2-b_{1}^2-b_{2}^2-b_{3}^2+b_{12}^2+b_{13}^2+b_{23}^2-b_{123}^2$\; $b_{I}= 2 b_{3} b_{12}-2 b_{2} b_{13}+2 b_{1} b_{23}-2 b_{0}
  b_{123}$\; $D=b_{S}^2+b_{I}^2$
\vskip 2pt
  {\scriptsize\tcc{Compute ($t_i,T_i$) pairs for $i=1$ and $2$, where $i=(1,2)\equiv(+,-)$}} \uIf{$-b_{S}+\sqrt{D}>0$}{$t_{1,2}=\frac{1}{4} \Bigl(b_{123} \pm
\frac{1}{\sqrt{2}}\sqrt{-b_{S}+\sqrt{D}}\Bigr),\
    T_{1,2}=\frac{1}{4}\Bigl(\frac{\pm b_{I}}{\sqrt{2}\sqrt{-b_{S}+\sqrt{D}}}-b_{0}\Bigr)$\;}
    \uElseIf{$-b_{S}+\sqrt{D}=0\ \mathrm{and}\ b_{S}>0$}{$t_{1,2}= \frac{1}{4}b_{123}, \
    T_{1,2}= \frac{1}{4} \bigl(\pm\sqrt{b_{S}}-b_{0}\bigr)${\scriptsize\tcc*[r]{root degenerate case}}}
    \Else{\Return{$\m{A}\leftarrow\emptyset$}{\scriptsize\tcc*[r]{no root}}}
  \vskip 2pt
 {\scriptsize\tcc{For each $(t_i,T_i)$  find corresponding $(+s_i,+S_i)$ and $(-s_i,-S_i)$}}
  \ForEach{$(t_i,T_i)$}{$
  s_{i} =\pm\sqrt{-T_i+\sqrt{T_i^2+ t_i^2}},\qquad S_{i}
  =\pm\frac{t_i}{\sqrt{-T_i+\sqrt{T_i^2+ t_i^2}}}$\;
  {\scriptsize\tcc{Value $i=1$ corresponds to sign plus and $i=2$ to sign minus}}
  }
%
 %
{\scriptsize\tcc{\!For each $(s,S)\to(s_i,S_i)$ pair find $(v_{i_k},V_{i_k})$ for $i=1,2$ and $k=1,2,3$\!\!}}
  \ForEach{$(s_i,S_i)$}{\eIf{$s^2+S^2\to s_i^2+S_i^2\neq 0$}{\tiny$
  \begin{aligned}
    v_{i_1}=\frac{b_1s_i+b_{23}S_i}{2(s_i^2+S_i^2)},\quad&v_{i_2}=\frac{b_{2}s_i-b_{13}S_i}{2(s_i^2+S_i^2)},\quad& v_{i_3}=\frac{b_3s_i+b_{12}S_i}{2(s_i^2+S_i^2)}\\
    V_{i_1}=\frac{b_{23}s_i-b_{1}S_i}{2(s_i^2+S_i^2)},\quad& V_{i_2}=-\frac{b_{13}s_i+b_{2}S_i}{2(s_i^2+S_i^2)},\quad& V_{i_3}=\frac{b_{12}s_i-b_{3}S_i}{2(s_i^2+S_i^2)}
\end{aligned}\;
  $\normalsize\newline
  \Return{$\m{A}_i\leftarrow (s_i,S_i, v_{i_k},V_{i_k})$ }{\scriptsize\tcc*[r]{isolated root coefficients}}}{
    \eIf{$b_1=b_2=b_3=b_{12}=b_{13}=$ $b_{23}=0${\scriptsize\tcc*[r]{$s^2+S^2=0$ case}}}{\scriptsize For a single pair $(v_{i_k},V_{i_j})$
    solve equations $\begin{cases}b_0=\bv^2-\bV^2,\\ b_{123}=2(\bv\cdot\bV)\end{cases}$\!\!\!.\; {\scriptsize The remaining pairs give free parameters.}\;{\normalsize
    \Return{$\m{A}_i\leftarrow (s_i,S_i, v_{i_k},V_{i_j})$ }}{\scriptsize\tcc*[r]{continuum of roots}}}{
    \Return{$\m{A}\leftarrow\emptyset$ }{\scriptsize\tcc*[r]{no roots}}
      }
  }}
  }\label{AlgForSqrt30} \caption{Algorithm for isolated roots of MV in real $\cl{3}{0}$}
\end{algorithm}

Roots: $\m{A}_{i_k}=\big((s_i+v_{i_1}\e{1}+v_{i_2}\e{2}+v_{i_3}\e{3})+
 (S_i+V_{i_1}\e{1}+V_{i_2}\e{2}+V_{i_3}\e{3})\e{123}\big)$


\bibliographystyle{REPORT}

\bibliography{SqrtRootMV}

\end{document}